\documentclass[12pt]{article}
\def\PARstart#1#2{\begingroup\def\par{\endgraf\endgroup\lineskiplimit=0pt}
    \setbox2=\hbox{\uppercase{#2} }\newdimen\tmpht \tmpht \ht2
    \advance\tmpht by \baselineskip\font\hhuge=cmr10 at \tmpht
    \setbox1=\hbox{{\hhuge #1}}
    \count7=\tmpht \count8=\ht1\divide\count8 by 1000 \divide\count7
by\count8
    \tmpht=.001\tmpht\multiply\tmpht by \count7\font\hhuge=cmr10 at \tmpht
    \setbox1=\hbox{{\hhuge #1}} \noindent \hangindent1.05\wd1
    \hangafter=-2 {\hskip-\hangindent \lower1\ht1\hbox{\raise1.0\ht2\copy1}%
    \kern-0\wd1}\copy2\lineskiplimit=-1000pt}

\if@technote\def\PARstart#1#2{#1#2}\fi     
\if@draftversion\def\PARstart#1#2{#1#2}\fi 

\def\keywords{\vspace{-.3em}
    \if@twocolumn
      \small\it Keywords\/\bf---$\!$%
    \else
      \begin{center}\small\bf Keywords\end{center}\quotation\small
    \fi}
\def\endkeywords{\vspace{0.6em}\par\if@twocolumn\else\endquotation\fi
    \normalsize\rm}

\def\BibTeX{{\rm B\kern-.05em{\sc i\kern-.025em b}\kern-.08em
    T\kern-.1667em\lower.7ex\hbox{E}\kern-.125emX}}
\usepackage{epsf}
\usepackage{graphics}
\usepackage{color}
\usepackage{psfig}

\def\IR{{\hbox{{\rm I}\kern-.2em\hbox{\rm R}}}}
\def\IH{{\hbox{{\rm I}\kern-.2em\hbox{\rm H}}}}
\def\IC{{\ \hbox{{\rm I}\kern-.6em\hbox{\bf C}}}}
\def\IZ{{\hbox{{\rm Z}\kern-.4em\hbox{\rm Z}}}}

\newcommand{\beq}{\begin{equation}}
\newcommand{\be}{\begin{equation}}
\newcommand{\eeq}{\end{equation}}
\newcommand{\ee}{\end{equation}}
\newcommand{\bea}{\par \begin{eqnarray}}
\newcommand{\eea}{\end{eqnarray}\par\par \noindent }
\newcommand{\bean}{\begin{eqnarray*}}
\newcommand{\eean}{\end{eqnarray*}}
\newcommand{\ba}{\beq\begin{array}{lll} }
\newcommand{\ea}{\end{array}\eeq}

\def\IC{ {\rm l\hspace{-1.2ex}C} }    
\def\IZ{{\hbox{{\rm Z}\kern-.4em\hbox{\rm Z}}}}
\def\IR{{\hbox{{\rm I}\kern-.2em\hbox{\rm R}}}}

\begin{document}

\title{A PIM-aided Kalman Filter for  GPS Tomography of the Ionospheric Electron Content}
\author{G. Ruffini, L. Cucurull, A. Flores, A. Rius}



\maketitle
\vspace{-1cm}

\begin{center}
Institut d'Estudis Espacials de Catalunya, CSIC Research Unit\\
Edif. Nexus-204, Gran Capit\`a, 2-4, 08034 Barcelona, Spain\\
Phone: +34 93 280 20 88   Fax:   +34 93 280 63 95 \\
e-mail: ruffini@ieec.fcr.es, http://www.ieec.fcr.es/earthsc-gb.html 
\end{center}

\begin{abstract} \normalsize\renewcommand{\baselinestretch}{1.5} \tiny \normalsize
 We develop the formalism for  a PIM-based functional for  stochastic tomography  with a Kalman filter, in which
  the inversion problem associated with four-dimensional ionospheric stochastic tomography is regularized. 
For consistency, GPS data is used to
select dynamically the best PIM parameters, in a 3DVAR fashion. We demonstrate the ingestion of GPS (IGS and GPS/MET) data into a parameterized ionospheric model, used to select the set of parameters that minimize a suitable cost functional. The resulting PIM-fitted model is compared to  direct 3D voxel tomography. We  demonstrate the value of 
this method  analyzing  IGS and GPS/MET GPS data, and present our results in terms of a 4D model
 of  the ionospheric electronic density.

\end{abstract}

\vspace{3cm} {\em \hspace{1cm} Submitted to Physics and Chemistry of the Earth}


\clearpage
\renewcommand{\baselinestretch}{1.5} \tiny \normalsize

\section{Introduction}
\PARstart{I}{n}  previous work \cite{ruffini98a, ruffini98b,ruffini97}, we  analyzed 
GPS data to extract information about
 the ionospheric electron density distribution.  We can think of this distribution as a field
 in space-time which we  try to represent using the information provided by the data. 
 Since the ionosphere produces
delays in the phase and group propagation of radio waves, 
having an accurate description of the electron content in the ionosphere is essential
to any endeavor that uses radio wave propagation (such as  tracking and navigating). In
 this paper we describe a novel parameterized tomographic technique to perform  ionospheric 
imaging  using Global Positioning System signal delay information. 

Climatological models of the ionosphere have existed for a while now, but it is only recently that they 
have been used to complement other sources of data, such as GPS, in the inversion process.
For instance, one can use input from a climatological model such as  PIM \cite{daniell}  to complement GPS data in 
the inversion process, and to compare the results to other data \cite{Manucci97}.  The parameters controlling 
the model are input directly, however, and are not estimated themselves. One could reason, however, that if the models 
were good enough they could used to infer these parameters given other sources of data, such as 
GPS ionospheric delay data. The resulting ``best-fit'' parameters should be related to 
the ones one can obtain by independent means.

Let us give a brief introduction to ionospheric tomography (more details can be found in 
\cite{ruffini98a, ruffini98b,ruffini97}).
Let $\rho(r,\theta,\phi,t)$ be the function that describes the  electron density in some region
of space ($r,\theta,\phi$ are spherical coordinates) at some time $t$.  We can  rewrite it as  
\beq
\rho(r,\theta,\phi,t)= \sum_J  a_J(t) \,  \Psi_J (r,\theta,\phi),
\eeq
where the functions $\Psi_J(r,\theta,\phi)$ can be any set of  basis
functions we like. 
The goal in  the inverse problem is to find the coefficients  $a_J(t)$. 
In the case of GPS ionospheric tomography we  use
 the information provided by the GPS ionospheric delay data along the satellite-receiver  rays
 $l_i$ to  obtain a set of 
equations, 
\beq\label{eq:main}
y_i = \int_{ l_i} d{l}\, \rho(r,\theta,\phi,t)= \sum_J  a_J(t) \int_{ l_i} d{l}\, \Psi_J(r,\theta,\phi),
\eeq
one for each ray $l_i$. Here   $y_i$ is the observed quantity. 
 This is a set of  linear equations of the form $A\,  x =y$,
where the components of the vector $x$ are the unknown coefficients $a_J(t)$. 
 Assume that some cut-off in the basis function expansion is used and, therefore,
that the $x$-space is $N$-dimensional. Let the $y$-space be  $M$-dimensional 
($M$  is thus the number of data points).
Since this system of equations may not have a solution 
 we seek to minimize the functional $\chi^2(x)$, where
(assuming uncorrelated observations of equal variance)
\beq
\chi^2 (x)= (y -A\, x)^T \cdot (y -A\, x).
\eeq
 In practice we  find that although the number of equations
 is much greater
 than the 
number of unknowns,   the unknowns, i.e., the array $x$,  are not completely fixed by the
 data. A way to restrict the solution space is to add some  a priori constraints to the problem,
and this can be implemented using   the Lagrange multiplier method.  Here we propose using a 
climatological model, such as PIM, to fill the gaps in the data and ``smooth'' the solution. 
  But in order to use a climatological model one must provide the necessary input parameters. It is certainly
 possible to use for these parameters  values provided by experimental sources of data (e.g., the solar flux is a measurable quantity). As was mentioned above, such techniques have already been used \cite{Manucci97}. But another way to proceed is to use the GPS 
data itself, together with the model, to fix these parameters.  This is especially important if it is suspected that the model parameters are not truly physical. If nothing else, 
this is an interesting exercise that will test the validity of the model.    

A  climatological mode, such as PIM, maps the value of a set of parameters, $\lambda_i$, to the
space $\{x\}$. Just as is done in variational weather modeling, we can picture minimizing the
cost functional
\beq
 J(\lambda_i)=\sum_j  \left( O^{exp}_j-O[{x}(\lambda_i)]_j \right)^2,
\eeq
where  {$O_j^{exp} $}   are the observables and  {$O[{x}(\lambda_i)]_j$} the modeled observables, in our case the slant delays produced by the ionospheric electrons.
 If we think of the climatological model image as  the space spanned by a set of 
empirical orthogonal functions (which is the case in PIM), 
we see that this approach is just as the one described before,
 in the sense that a finite basis set is used to fit the data and represent the solution. 
 What a  model like PIM  does is to provide us 
with a set
of empirically or theoretically optimized basis functions to represent the ionospheric electron content. 

 \section{PIM-aided Kalman Filtering}
Kalman filtering is a very useful technique when  dealing with a dynamic process in which  
data is available at different times.  It is a natural way to enforce smoothness under time
 evolution, and is
 especially useful in the case of ionospheric stochastic tomography, when the ``holes'' in the
 information that 
we have at a given time (because of the particular spatial distribution of the 
 GPS constellation  and the receptor grid) 
may be ``plugged'' by the data from  
previous and future
 measurements.
Indeed, in a Kalman filter we use the  information contained in  a  solution to the inversion 
problem to estimate the next solution in the iteration process. 
In the study of the ionosphere, for example, we break
the continuous flow of satellite delay data into blocks  of a few hours, and simply model ionospheric 
 dynamics
 by a random walk \cite{Herring}. We can
then process the data at a given point in the iteration  by  asking that, to some extent, the solution 
   be similar to the one in the previous iteration, depending on how much 
 confidence we have in that previous solution,  and on how much 
 we expect the dynamics to have changed things from one solution to the next.  Here we complement this step 
by using the previous solution in the iteration process to fit a PIM model to the data. 
 In other words, if $x_n$ and $C_{n}$ are the solution and
the  covariance matrix at epoch  $n$, we first determine a minimum squares PIM fit. 
 Let $A$ be the observation matrix (which we know how to compute, given a grid). 
Then we  minimize the cost functional
\beq
J(\lambda)= \left(y-A \cdot x^{PIM}(\lambda)\right)^2, 
\eeq
and this will determine the PIM parameters $\lambda^i$, and the resulting image, $x^{PIM}_n(\lambda)$ and  covariance matrix for the voxel image, $C^{PIM}_n$. This matrix  is related to  the covariance matrix for the PIM parameters,
\beq
C^{-1} = \nabla_\lambda\nabla_{\lambda'} J ,
\eeq
and is given by
\beq
C^{PIM}_n= \left(\nabla_\lambda x^i(\lambda)  \left( \nabla_\lambda\nabla_{\lambda'} J \right)^{-1}\nabla_\lambda x^j(\lambda')\right)^{-1} .
\eeq 
We will  not worry too much about it for now, since it may be hard to compute these PIM
  derivatives. We will instead use an {\it ad hoc} covariance matrix, with the property that it will fill the holes in the data without affecting too much
 the solution where the data already provides some information (as is done in  \cite{Manucci97}).

Since the extremization equation for this functional is not linear and we could not easily compute derivatives we have
chosen to
 minimize this functional using the Powell  algorithm
(see \cite{recipes}, for example).

 Now, at  epoch $n+1$ we are to minimize
\beq
{\cal K}_{n+1}=  \chi_{n+1}^2(x_{n+1})
+   \left( x_{n+1}-x^{PIM}_{n}(\lambda) \right)^T     
 \left(C^{PIM}_{n+1}+\delta^2 \right)^{-1}  
\left(x_{n+1}-x^{PIM}_{n}(\lambda)\right)
\eeq
with respect to $x_{n+1}$. The parameter 
  $\delta$ (which will in general be a diagonal $N\times N$ matrix) models the random walk away from
 the previous solution, and if of the form $\delta^2= \alpha \cdot t$. 
Minimization yields
\beq
x_{n+1} = \left[S_{n+1}+ \left(C_{n}^{PIM}+\delta^2 \right)^{-1}\right]^{-1}
\left(   A^T_{n+1}y_{n+1}  + \left(C_{n}^{PIM}+\delta^2 \right)^{-1} x_{n}^{PIM}  \right) ,
\eeq
where $S_n=A_n^TA_n $, and $C_{n}^{-1}= S_n +  \left(C_{n-1}^{PIM}+\delta^2 \right)^{-1}$.
This  can be easily implemented in an algorithm.

\section{Ingesting GPS data into PIM versus using regular tomography}\label{GPS} 
Let us first summarize our goals: 
\begin{itemize}
\item To demonstrate the ingestion of GPS (IGS and GPS/MET) data into a parameterized ionospheric model, and to select the set of parameters that minimize a suitable cost functional.
\item To compare the model fit to direct 3D voxel tomography.
\item  To develop  a PIM-based functional for  stochastic tomography  with a Kalman filter, in which
  the inversion problem associated with four-dimensional ionospheric stochastic tomography is regularized. 
For consistency, GPS data is used to
select dynamically the best PIM parameters, in a 3DVAR fashion.  
\end {itemize}
GPS observables consist essentially of the delays experienced by the dual frequency
signals 
($f_1 = $1.57542 GHz  and $f_2=$1.22760 GHz)  transmitted from
the GPS constellation (25 satellites) and received at GPS receivers around the world and in orbit.
 Let  $L_i$ be the measured  total flight time
 in light-meters of  a ray  going from 
 a given GPS 
satellite to a receiver at the frequency $f_i$ (including
 instrumental biases),  and  $I=\int_{ray} {dl} \, \rho(x)$
 be the integrated  electron density along the ray  
(in electrons per square meter).
Then  $L_i$ is modeled  by  $L_i = D- I \, \alpha/f_i^2 + \tilde{c}_{sat} +\tilde{c}_{rec}$,
 where
$\alpha= 40.3 \, m^3/s^2$,  $D$ is the length of the ray, and  
$\tilde{c}_{sat} $ and $\tilde{c}_{rec} $ are the  
instrumental biases.
 In the 
present case we are interested in the frequency dependent part of the delay: 
${L}={L}_1-{L}_2$ (in meters). This  is the derived observable and is modeled by
 ($\gamma= 1.05 \times 10^{-17} $ $m^3$)
$
L= \gamma \, I  + c_{sat} + c_{rec}
$, 
 independent of $D$ (see \cite{ruffini98b} for more details).  For the purposes of PIM-fitting, the solutions
 for the bias constants from the previous iteration are used to ``fix'' the observables delays, so that
 only the electronic part of the delay remains. At this point we have not tried to estimate the bias 
constants within the PIM-fitting analysis, although this should be possible.  
See the Appendix A for details on our bias constant treatment.
 
GPS data has been  collected from  GPS/MET and a subset of the International GPS Service (IGS) Network, for 
the day  of February 23rd of  1997. 
 This particular day has been  chosen because of A/S is known to have been off. 
 Geomagnetic and solar
activity indices (as distributed by the US  National Geophysical Data Center) for that day indicate a mean {$K_p$ index of 2.3}, and  {$F_{10.7}=73$}.  

The raw data has been  pre-processed in order to obtain the observables
using the procedures described in \cite{ruffini98b}.
To describe the ionosphere we use   {five} geocentric spherical  layers
 beginning at {50 km}  above 
the mean surface (6350 km) of the Earth and extending {1300 km}. Each layer consists then  of two  hundred  
 voxels of dimensions {18$^o$} in latitude, times   {18$^o$} in longitude, times   150 km of height for the first 4 layers.

The
unknowns here  consist of the electron densities at each of these voxels, 
plus the unknowns corresponding to
the transmitter and receiver constant delays.   
 These are estimated and used to correct the data prior to PIM-fitting.  For a particular block, a minimum was found at  $F_{10.7}=52$ and $ K_p=0$. Thus, we see that these parameters should not be taken as physical quantities but just as parameters in the model. The PIM fit had a reasonable quality (40 cm standard deviation).  Using the parameters estimated form observation ($F_{10.7}=73$ and $ K_p=2.3$) yields a standard deviation of 45 cm (they are far from the minimum). This is expected, as it is known that PIM tends to overestimate TECs ({\em Rob Daniell, private communication}).

\section{Summary, Conclusions}

In this paper we have summarized our efforts to use climatological models in tomographic analysis of GPS data. 
This is a more natural thing to try than one may think at first.  After all, climatological models such 
as PIM are essentially the result of performing  Empirical Orthogonal Function analysis using empirical 
or theoretical data, and in a way this
 is exactly what one would like to do in tomography:  the basis functions used to span the space
 of possible solutions should be adapted to the field one is trying to map.  Basis sets such as
 wavelets are a step in this direction, but they are optimized to attack more general problems, 
where certain characteristics of the field one is studying are known.  Here we can refine the basis
 set even more, given the theoretical and 
experimental knowledge that we  already posses about the ionosphere.  We have seen that the parameters in the model are not really physical, and we conclude that it is necessary to perform such parameter fits prior using the model estimates in the Kalman filter. Future efforts should be directed towards
 the development of
 more refined parameterized models.  The ingestion of GPS data into this type of model has been demonstrated here.

\section*{Appendix A}
Here we show how to take out the constants from the analysis.  Let $x$ denote the array solution, in which the first $n$ entries correspond to the voxel unknowns, and thereafter to the bias constants. Let us rewrite 
$
x=x_{vox}+x_c,
$
where $x_{vox}$ is an array with zeros after the $n$th entry, and $x_c$ has zeros until after the $n$th entry. Now,
\bea
\chi^2 (x)&=& (y -A\, x)^T \cdot (y -A\, x) \nonumber\\
          &=& y^T y + x_c^T A^T A x_c - 2x_c^TA^T y +  x_{vox}^T A^T A x_{vox} + 2  x_{vox}^T\left( -A^T y+A^T A x_c \right).
\eea
Hence, if we wish to fix $x_c$, all that is needed is to modify $A^T y \rightarrow A^T (y- A x_c)=\left(A^Ty\right)_{corr}$, and proceed without estimating the constants.  Since $x_{vox}$ is an array with zeros after the $n$th entry, only the first $n$ terms of  $\left(A^Ty\right)_{corr}$ are needed. The terms $y^T y + x_c^T A^T A x_c- 2x_c^TA^T y $ are constants and do not affect the minimization solution.  Hence we see that, up to irrelevant constant terms, the minimization problem is the same as without constants, but with a modified $A^Ty$ term.
\renewcommand{\baselinestretch}{1.0} \tiny

\newpage

\begin{figure}[h!] 
\epsfysize=10cm 
\hspace{-0.5cm}  
{\epsffile{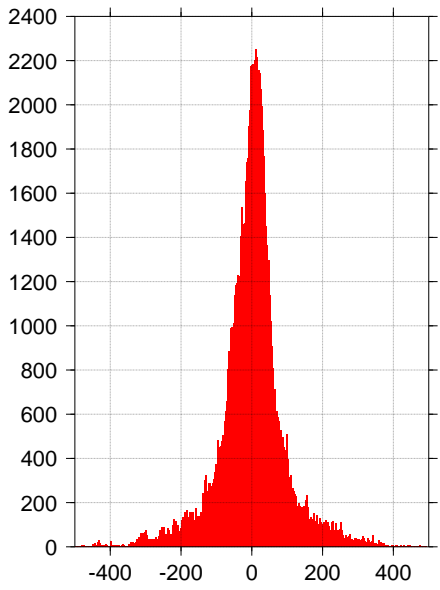}   }\epsfysize=10cm { \epsffile{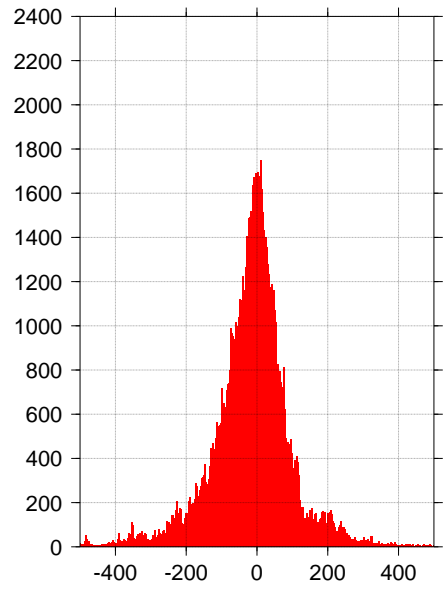}   }
\epsfysize=10cm { \epsffile{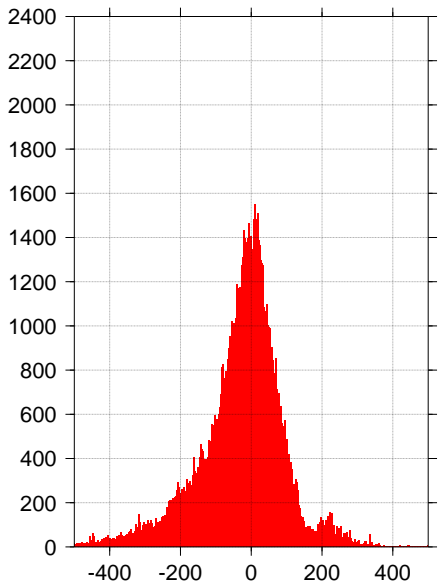}   }
\caption{\label{tomo_hist.ps} Left: Tomogrphic residual histogram. Standard deviation is 30 cm. Middle: PIM-fit residuals (at $F_{10.7}=52$ and $ K_p=0$). Standard deviation is 40 cm. Right: PIM-fit residuals (at $F_{10.7}=73$ and $ K_p=2.3$). Standard deviation is 45 cm. }
\end{figure}

\clearpage
\vspace{-3cm}

 \hspace{4cm}

\begin{figure}[h!] \hspace{2cm}\parbox[b]{10cm}{
\epsfysize=20cm 
\hspace{-4cm} 
{\epsffile{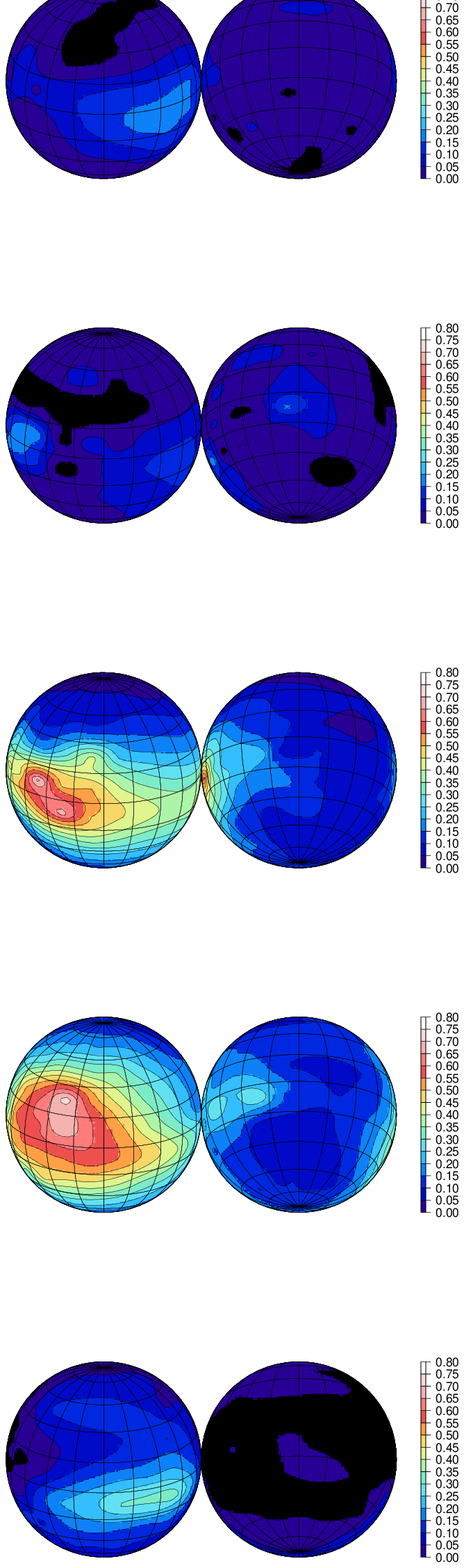}   } }\hspace{-8.5cm} \parbox[b]{8cm}{
\epsfysize=20cm { \epsffile{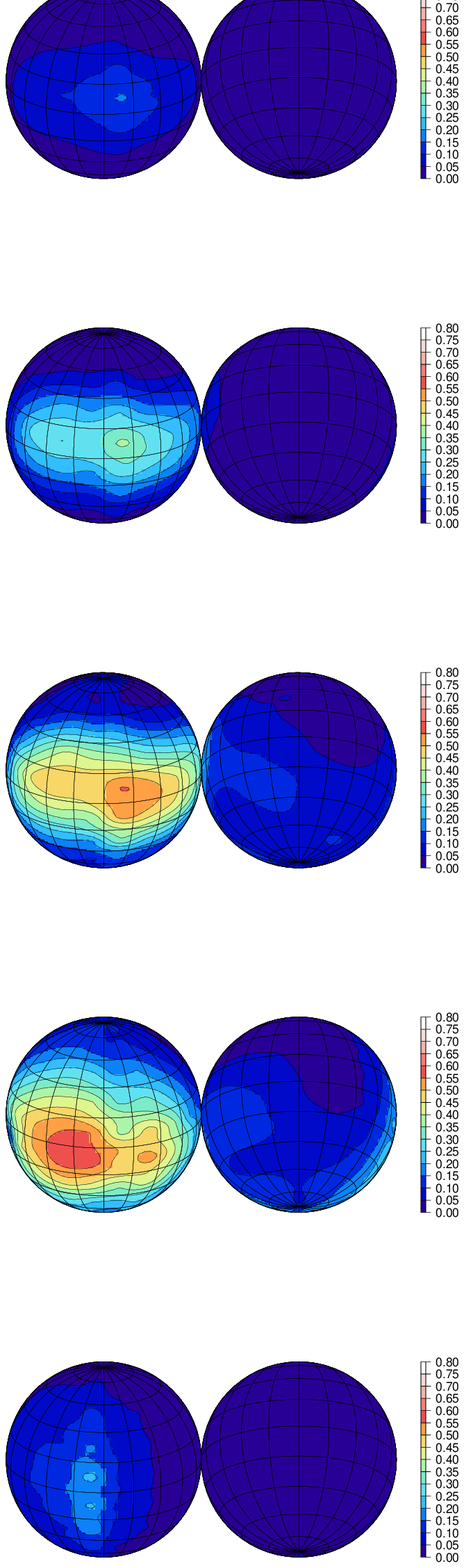}   }   }\vspace{-3cm}

\vfill
 { \noindent {\bf Figure 2:} Tomographic solution (left column) and PIM-fit solution (right column), layer by layer and from bottom up, 6400-6550, 6550-6700, 6700-6850, 6850-7000, 7000-7700 km from center of Earth.  Electronic density units are  Tera electrons ($10^{12}$) per cubic meter.} 

\end{figure} 

\end{document}